# Regularization of Vertical-Cavity Surface-Emitting Laser's emission by periodic non-Hermitian potentials

W. W. Ahmed,[1*] R. Herrero,[2] M. Botey,[2] Y. Wu,[1] K. Staliunas[2,3]

[1]Division of Computer, Electrical and Mathematical Sciences and Engineering, King Abdullah University of Science and Technology (KAUST), Thuwal, 23955-6900, Saudi Arabia
[2]Departament de Física, Universitat Politècnica de Catalunya (UPC), Colom 11, E-08222 Terrassa, Barcelona, Spain
[3]Institució Catalana de Recerca i Estudis Avançats (ICREA), Passeig Lluís Companys 23, E-08010, Barcelona, Spain
*Corresponding author: waqas.waseem291@gmail.com

**We propose a novel physical mechanism based on periodic non-Hermitian potentials to efficiently control the complex spatial dynamics of broad-area lasers, particularly in Vertical-Cavity Surface-Emitting Lasers (VCSELs), achieving a stable emission of maximum brightness. Radially dephased periodic refractive index and gain-loss modulations accumulate the generated light from the entire active layer and concentrate it around the structure axis to emit narrow, bright beams. The effect is due to asymmetric-inward radial coupling between transverse modes, for particular phase differences of the refractive index and gain-loss modulations. Light is confined into a central beam with large intensity opening the path to design compact, bright and efficient broad-area light sources. We perform a comprehensive analysis to explore the maximum central intensity enhancement and concentration regimes. The study reveals that the optimum schemes are those holding unidirectional inward coupling but not fulfilling a perfect local PT-symmetry.**

Semiconductor lasers are compact and efficient coherent light sources used for applications ranging from data processing to optical communications. Among such lasers are Vertical-Cavity Surface-Emitting Lasers (VCSELs), where the beam emission direction is perpendicular to the active region [1]. Semiconductor micro-lasers, including VCSELs, are generally unstable, especially in broad emission area regimes: random fluctuations and spatiotemporal instabilities, arising from modulation instability, degrade the spatial beam quality and laser coherence [2-4]. This intrinsic instability stems from the self-focusing nonlinearity and the lack of an intrinsic transverse mode control that gives rise to different filamentation regimes. In addition, carrier hole burning leads to irregular pulsating patterns [5]. Common techniques to control the complex dynamics of semiconductor lasers rely on optical feedback and injection [6-10]. Yet feedback-based beam control strategies are only applicable to the specific configurations, and severely reduce the compactness of the design. For instance, VECSELs with a large external cavity use a spherical mirror to induce single transverse mode operation [11], which also reduces the power conversion efficiency since the spherical mirror limits the effective active region. Therefore, there is a need for a more general physical mechanism to control the complex spatial dynamics in semiconductor lasers, for a high spatial quality emission.

In recent years, non-Hermitian spatially modulated materials have provided a flexible platform to manipulate lightwave dynamics. The simultaneous refractive index and pump modulations have already shown the capability to suppress spatial instabilities in nonlinear optical systems, particularly in broad-area semiconductor (BAS) and VECSEL devices [12-14]. A particularly remarkable class of such materials is those with Parity-Time (PT-) symmetry [15]. In such systems, the complex refractive index representing refractive index (real part) and gain-loss (imaginary part), fulfills: $n(x)=n^*(-x)$ i.e., the real part of the refractive index is symmetric while the imaginary part is antisymmetric in space [16]. PT-symmetric realizations in optics lead to unconventional beam dynamics [17-19]. One of the most interesting features of such materials is the unidirectional light transport arising from the unidirectional coupling at the phase transition point. In periodic PT-symmetric media, where index and gain-loss modulations are dephased by a quarter of wavenumber of the modulation, the exceptional point (maximal asymmetry) occurs when the gain and loss modulation amplitudes are balanced [20]. PT-symmetric optical potentials uncover novel physical effects such as unidirectional invisibility [21], power oscillations [22], coherent perfect absorption [23], single mode lasing [24, 25] among others. Recently, a new class of so-called *local* PT-symmetric potentials was proposed to manipulate the electromagnetic field flows in any desired configuration [26, 27], where the locally symmetry broken potentials ensure the local directions of the field flows.

In this letter, we propose to apply non-Hermitian potentials to control the spatiotemporal dynamics in VCSELs. The unidirectional inward radial mode coupling due to appropriate non-Hermitian potentials is expected to collect the energy from the entire active region and concentrate it around the center to form localized beams. While typical VCSEL's emission exhibits complex and extended spatiotemporal dynamics, as shown in Fig. 1(a), the proposed VCSEL scheme aims at improving its emitted light by concentrating it into a narrow and bright beam as shown in Fig. 1(b). Such narrow beam emission is expected to enhance the performance and effectiveness of the system, without losing its compactness, which could be important for a large variety of practical applications.

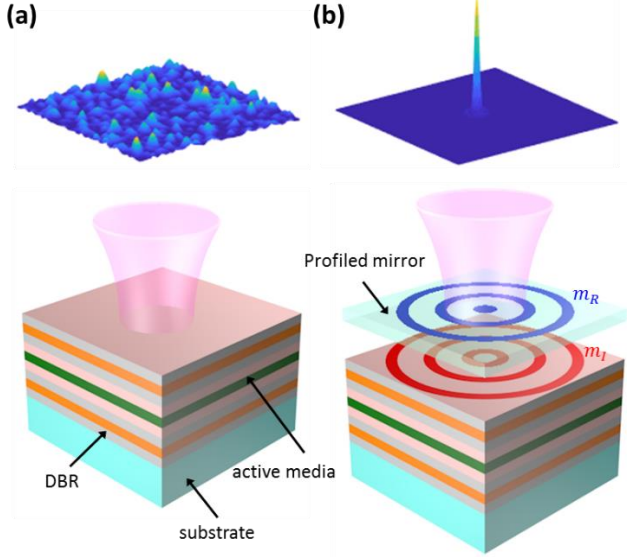

Fig. 1. VCSEL's emission, schematic illustration: (a) complex irregular spatial pattern emitted from a conventional broad-area VCSEL, (b) bright and narrow beam emission from the modified VCSEL with a non-Hermitian concentric configuration, where $m_R$ and $m_I$ denote the real (refractive index) and imaginary (gain-loss) parts of the complex refractive index.

The spatiotemporal evolution of the slowly varying intracavity field, E and carrier density, D, in VCSELs can be described by the system of coupled rate equations [28]:

$$\partial_t E(r,t) = -[1+i\theta+2C(i\alpha-1)(N-1)]E + i\nabla_\perp^2 E + iV(r)E$$

$$\partial_t N(r,t) = -\gamma[N - I_p + |E|^2 (N-1)] + \gamma d\nabla_\perp^2 N \qquad (1)$$

Here time is scaled to the cavity photon lifetime, $\tau_p = 2L/v(1-R)$ $R$ being is the reflection coefficient of mirrors, $v$ is the velocity of light, and $L$ is the length of the laser cavity. The transverse coordinate $r$ is scaled to the diffraction length of the cavity $\sqrt{\lambda L/2\pi(1-R)}$, where $\lambda$ is the emission wavelength. The transverse Laplacian terms $\nabla_\perp^2$, account for field diffraction and carriers' diffusion, $\theta$ represents the cavity detuning parameter, $\gamma$ is the carrier decay rate normalized to the photon relaxation rate, $\alpha$ is the Henry (linewidth enhancement) factor of the semiconductor, C is the pump parameter, d is the carrier diffusion coefficient, and $I_p$ is the normalized pump current. Finally, $V(r)$ represents a non-Hermitian potential, which can be expressed in axisymmetric-harmonic form: $V(r) = m_R \cos(q_r r) - i m_I \cos(q_r r - \phi_n)$ with $m_R$ and $m_I$ being the amplitudes of the real (refractive index) and the imaginary (gain-loss) modulations, with the small-scale spatial modulation wavenumber: $q_r$. The relative phase shift between real and imaginary modulations is denoted by $\phi_n$.

We start from a simplified VCSELs modulated in one transverse dimension (1D). We numerically integrate Eq. (1) for: $V(x) = m_R \cos(|x|) - i m_I \cos(|x| - \phi_n)$, using a split-step method. The introduction of the non-Hermitian potential aims at directing the light, generated from the whole cavity, towards the center of the device for an extraordinary intensity enhancement and concentration around the center. Note the presence of moduli of the coordinate, which ensures the symmetry of the potential with respect to $x$. The performance of the proposed modulated VCSEL is evaluated by two parameters: the enhancement of central intensity, $I(x=0)$, and the concentration factor $C_f = I(x=0)/\langle I(x) \rangle$. We explore the parameter space $(m_I, \phi_n)$ for a fixed value of $m_R$ and summarize the results in Fig. 2. In this 1D scheme, the local axial PT-symmetric potential imposes a unidirectional mode coupling towards the symmetry axis, at $x=0$ [26]. The inward mode coupling is expected to accumulate the field for balanced index and gain-loss modulations (and $\phi_n = 90°$). Figures 2(a) and 2(b) depict the central enhancement and axial concentration maps, respectively in parameter space $(m_I, \phi_n)$. As expected, when the mode coupling is strongly outwards (around $\phi_n \sim 270°$) no concentration is found. We observe, however, that both the central intensity and the concentration factor are maximized not exactly at the phase of PT-symmetry, i.e. for $\phi_n = 90°$ but with a slight deviation, $\phi_n \sim 80°$. This may be attributed to the fact that for $\phi_n$ values slightly smaller than 90° the gain area around the center is enlarged. Note that while central intensity grows radially as shown in Fig. 2(a) (increasing the gain-loss modulation amplitude), Fig. 2(b) reveals an island of field concentration in which different sets of parameters (gain-loss modulation amplitude and the phase shift between the index and gain-loss modulations) may lead to an intense localized beam emission.

In order to clarify this apparent difference between Figs. 2(a) and 2(b) and identify the possible operating regimes, we investigate the temporal dynamics of modulated VCSELs. The spatial and temporal dynamics of three representative points are presented in Fig. 2 (c,d,e). The laser dynamics exhibits two possible operating regimes i.e. stationary and oscillatory depending on the relative modulation amplitudes and phase. While Fig. 2(c) corresponds to perfect PT-symmetric scheme, intensity and field concentration are much higher in Fig. 2(d). Further increasing the gain-loss amplitude as in Fig. 2(e), the oscillatory regime appears since for higher gain intensity is restricted due to saturation. The insets on the left panels show the potential profiles. Thus, the system shows a stationary state in (c,d) and the pulsating temporal behavior in (e), see the temporal central intensity evolution on the right panels. In all cases, we observe relaxation oscillations inherent of Class-B dynamic regime in transients [14]. We note that the results persist for larger values of the carrier decay rate, i.e. regardless of the laser class dynamical limit. Similar results are found for Class-B lasers with $\gamma \sim 0.1$, the intermediate regime and Class-A lasers ($\gamma \gg 1$), signifying the robustness of the proposed field concentration effect.

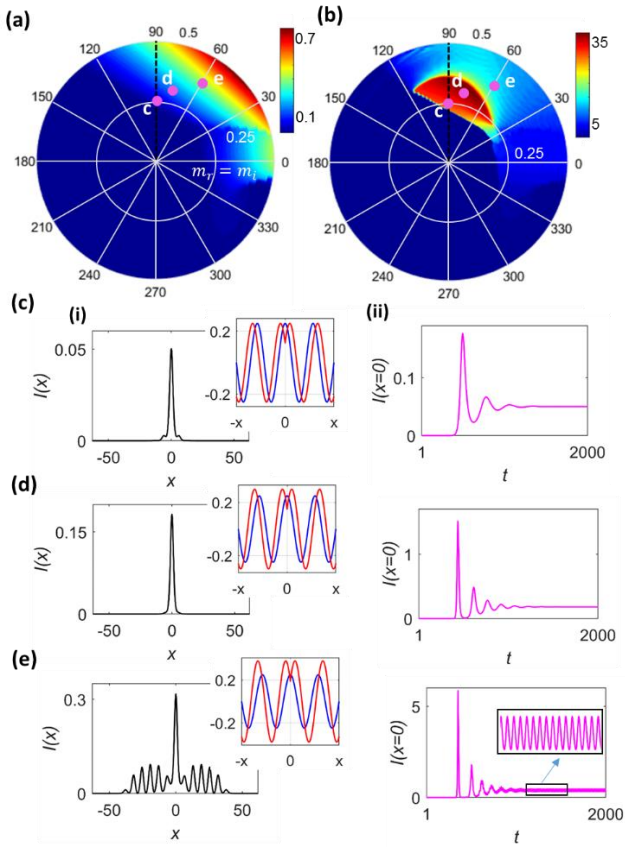

Fig. 2. 1D modulated VCSELs. Central intensity map (a) and concentration factor map calculated after long time evolution (b), for $m_R = 0.25$, $m_I = [0, 0.5]$ and $\phi_n = [0, 360°]$. The spatial intensity profiles and time evolution of the central intensity are shown in the columns (i) and (ii) for representative points (c,d,e). The insets on the left panels represent the real (blue) and gain-loss(red) modulations of the non-Hermitian potential for each point. Other system parameters are: $\alpha=2$, $C=0.6$, $\theta=-2$, $\gamma=0.01$, $I_P=1.75$, and $d=0.05$.

In the above study, the considered potential fixed a maximum of the refractive index profile at the center and determined the corresponding phase of the gain modulation. For this case, we found a maximum enhancement of the central intensity for $\phi_n < 90°$. We study the role of the center character by introducing an additional phase in both modulations, referred as the central phase, $\phi_c$. The corresponding 1D non-Hermitian potential may then be rewritten as $V(x) = m_R \cos(|x| + \phi + \phi_n) - i m_I \cos(|x| + \phi)$ with $\phi = \phi_c - \phi_n$ and explore the parameter space $(\phi, \phi_n)$ for balanced modulation amplitudes, $m_R = m_I = 0.25$. The results are plotted in Fig. 3. Figures 3(a) and 3(b) depict the central enhancement and concentration factor, respectively. We observe two islands of high peak intensity in Fig. 3(a) for different sets of central and non-Hermitian phases where the dotted black curve delimits the stationary and oscillatory regimes in parameter space. We consider three representative points (c,d,e) to illustrate the spatial dynamics of the emitted beam. In (c) and (d), the central intensity is almost comparable, yet the spatial profile is broader in (d) since concentration is smaller. The index and gain profiles around the center are depicted in insets. The side lobes in the emitted beam develop when the gain is maximum at the center finally leading to a decrease in concentration and central intensity [see Fig. 3(e)]. We note that the phase, $\phi$, determines the width of the gain area around the center and is therefore related to the width of the envelope of the emitted beam. The smallest peak width of the localized beam takes place for $\phi = -60°$ where the gain area reaches the maximum width and for a given width of the localized beam, there is an optimal phase indicated by a solid black line in Figs. 3(a,b).

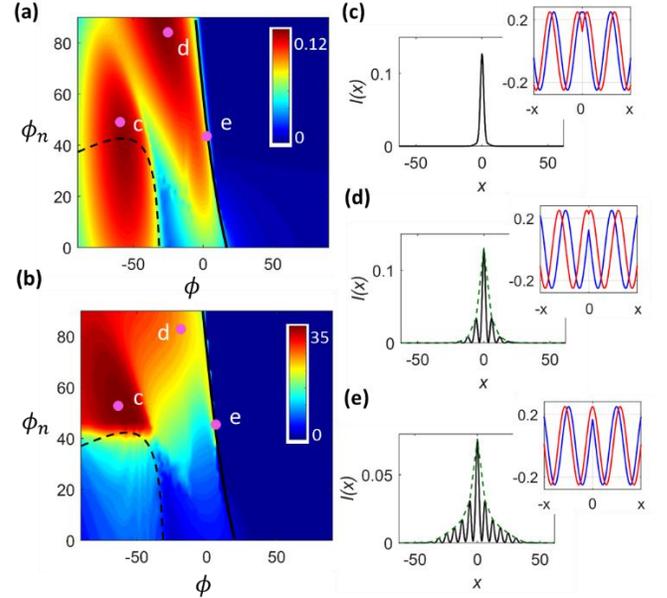

Fig. 3. Dynamics of 1D modulated VCSELs depending on the total phase central and non-Hermitian phase. Central intensity map (a), and concentration factor map calculated after a sufficiently long time (b), in parameter space $(\phi, \phi_n)$, for $m_R = m_I = 0.25$. The spatial intensity profiles of the stationary states are shown for the representative points (c,d,e). The other system parameters are the same as in Fig. 2.

Finally, we extend our analysis to the more realistic 2D modulated VCSEL with the potential: $V(r) = m_R \cos(r) - i m_I \cos(r - \phi_n)$. Analogously to the 1D study, we find that the intensity is concentrated towards the center due to a unidirectional radial inward coupling of the transverse modes. The results for the 2D system are presented in Fig. 4. Higher intensities and concentration factors are found in this case possibly due to a larger mode coupling and a larger gain area around the center, $r=0$, as compared to the 1D case. The optimum concentration regime [see Fig. 4(b)] occurs in this case also for slightly unbalanced relative amplitudes of the index and gain-loss modulations, for $\phi_n \sim 105°$. For this phase difference, the corresponding potential exhibits a small loss at the center, which prevails from saturation and induces a sharper beam by decreasing the energy of the side lobes. The intensity and the corresponding axial cross-sectional profiles for different parameter sets are shown in Figs. 4(c,d). We also present the corresponding transverse field flow, obtained as $F = i(E\nabla E^* - E^*\nabla E)$ in the corresponding insets. Indeed, the flow illustrates the expected radial inward wave flow arising from the unidirectional radial coupling, which leads to the field concentration around the center, $r=0$.

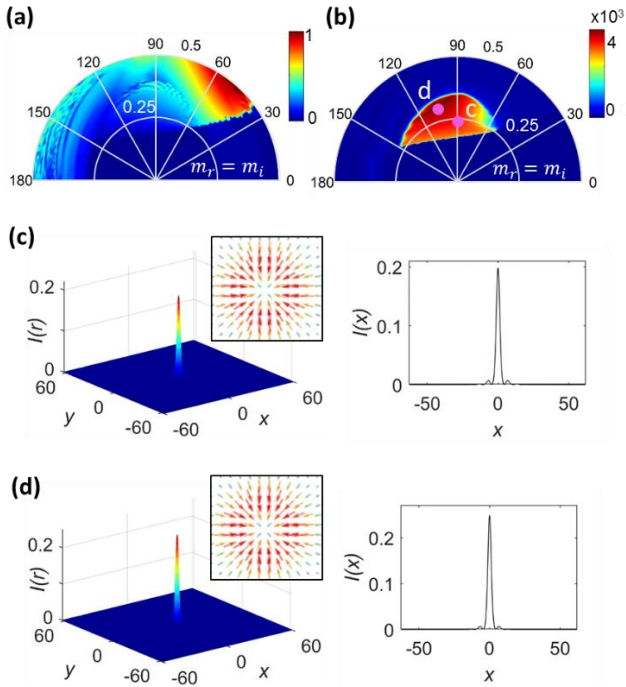

Fig. 4. 2D modulated VCSELs. Central intensity map (a), and concentration factor map after sufficient long evolution time (b), for $m_R = 0.25$, $m_I = [0, 0.5]$ and $\phi_n = [0, 180°]$. The spatial dynamics for two representative points is shown in (c, d) where the left panels depict the intensity emission profiles in the stationary state, and the right panels show the corresponding axial cross-sections. The insets illustrate the transverse field flow pattern around the center, $r=0$, that leads to intensity enhancement and concentration. All other parameters are the intensity enhancement and concentration. All other parameters are the same as in Fig. 2.

To conclude, we propose non-Hermitian potentials to regularize the complex spatial dynamics of VCSELs. The asymmetric inward coupling among the transverse modes concentrates the light around the center rending BAS lasers into bright and narrow-beam sources. We find the maximum central intensity and concentration regimes in the parameter space by exploring the modulation parameters. We observe such lasers can be operated in stationary or oscillatory regime depending on the relative amplitude and phase of the index and gain-loss modulations. The results indicate a significant intensity enhancement and concentration in the emitted beam when the coupling between transverse modes is inwards yet not fulfilling perfect local PT-symmetry. The 2D study uncovers rich possibilities for various configurations which could be extended beyond periodic non-Hermitian potentials assuming different random, quasiperiodic complex profiles of the background potential etc. Moreover, the proposed scheme is shown to be robust, compact and efficient and may be applicable to other BAS lasers and microlasers to improve their performance.

**Acknowledgments**: The work described in here is partially supported by King Abdullah University of Science and Technology (KAUST) Office of Sponsored Research (OSR) under Award No. OSR-2016-CRG5-2950 and KAUST Baseline Research Fund BAS/1/1626-01-01. K. Staliunas acknowledges the support of Spanish Ministerio de Economía y Competitividad (FIS2015-65998-C2-1-P) and European Union Horizon 2020 Framework EUROSTARS (E10524 HIP-Laser).